# Neuromorphic Mimicry Attacks: Exploiting Brain-Inspired Computing for Covert Cyber Intrusions

Hemanth Ravipati [1]

[1] hr48s@missouristate.edu

[1]Computer Science, Missouri State University, Texas, USA

*Abstract*— Neuromorphic computing, inspired by the human brain's neural architecture, is revolutionizing artificial intelligence and edge computing with its low-power, adaptive, and event-driven designs. However, these unique characteristics introduce novel cybersecurity risks. This paper proposes Neuromorphic Mimicry Attacks (NMAs), a groundbreaking class of threats that exploit the probabilistic and non-deterministic nature of neuromorphic chips to execute covert intrusions. By mimicking legitimate neural activity through techniques like synaptic weight tampering and sensory input poisoning, NMAs evade traditional intrusion detection systems, posing risks to applications such as autonomous vehicles, smart medical implants, and IoT networks. This research develops a theoretical framework for NMAs, evaluates their impact using a simulated neuromorphic chip dataset, and proposes countermeasures, including neural-specific anomaly detection and secure synaptic learning protocols. The findings underscore the critical need for tailored cybersecurity measures to protect brain-inspired computing, offering a pioneering exploration of this emerging threat landscape.

*Index Terms*— Neuromorphic computing, cybersecurity, mimicry attacks, brain-inspired systems, and intrusion detection.

## I. INTRODUCTION

Neuromorphic computing represents a paradigm shift from traditional von Neumann architectures to systems that mimic the brain's neural structure [1]. These chips feature low power consumption, event-driven processing, and adaptive learning, making them ideal for AI, edge computing, and IoT applications [2]. By 2025, neuromorphic systems are increasingly adopted in autonomous vehicles, smart implants, and decentralized IoT networks due to their efficiency in resource-limited environments [3]. However, their core traits—probabilistic behavior, non-deterministic processing, and synaptic weight dependency—introduce novel and unaddressed cybersecurity risks [4].

This research introduces Neuromorphic Mimicry Attacks (NMAs), a new class of cyber threats that exploit neuromorphic chips' biological-like behavior [5]. Unlike conventional attacks, NMAs manipulate neural dynamics to mimic legitimate activity, often bypassing existing intrusion detection systems (IDS) [6]. Key techniques include synaptic weight tampering and sensory input poisoning, which subtly alter system outputs without raising alarms [7]. Such attacks could misdirect autonomous navigation or interfere with real-time medical monitoring, leading to potentially catastrophic outcomes [8].

The motivation behind this study stems from the fast-growing integration of neuromorphic systems and the lack of security models tailored to their architecture [9]. While traditional security efforts focus on software exploits or hardware Trojans [10], neuromorphic systems demand distinct solutions due to their analog, decentralized design [11]. This work fills that gap by proposing a structured NMA framework, simulating attacks using a custom dataset, and developing countermeasures that align with neuromorphic processing principles [12]. The dataset, derived from a simulated neuromorphic chip, captures neural behavior under both normal and compromised conditions [13]. Unlike datasets such as CIC-IDS-2017 [14], which log network traffic, this dataset targets hardware-level neural interactions [15]. Data visualizations including bar graphs and tables highlight the effectiveness and stealth of NMAs, supporting the study's findings [16].

The primary objectives are: (1) to define and characterize NMAs, (2) to assess their impact, and (3) to propose specialized defensive strategies [17]. By achieving these, this paper aims to pioneer security research in neuromorphic computing and support its safe and reliable deployment [18][19].

## II. THEORETICAL BACKGROUND

Neuromorphic computing emulates the brain using spiking neural networks (SNNs) and event-driven processing for energy-efficient computation [20]. Unlike CPUs, neuromorphic chips operate asynchronously, firing neurons only upon stimulus [21]. This model, seen in IBM's TrueNorth and Intel's Loihi, enables real-time adaptability for autonomous systems and IoT applications [22]. However, the inherent probabilistic and analog behaviors introduce security challenges beyond conventional frameworks [23].

Neuromorphic Mimicry Attacks (NMAs) exploit the brain-like nature of these chips by manipulating synaptic weights and

4poisoning sensory inputs [24][25]. Synaptic weights control neuron connectivity and system learning [1], and minor alterations can bias outcomes without detection [2]. Sensory input poisoning introduces malicious signals resembling legitimate data, misleading system behavior [3]. NMAs align with stealth-based cybersecurity threats like advanced persistent threats (APTs) but target hardware-level vulnerabilities [4][5]. Neurologically, they mimic brain pathologies, such as neural misfiring, blending with valid spike activity [6][7]. A formal NMA model includes the attack vector (e.g., weight tampering), the target system (neuromorphic chip with SNN), and the impact mechanism (e.g., misclassification or latency shifts) [9][10][12].

Countermeasures must align with neuromorphic dynamics. Traditional intrusion detection systems (IDS) are ineffective in probabilistic contexts [14][15]. Alternatives include anomaly detection tailored to spike dynamics and secure learning protocols for synaptic integrity [16]. These approaches adapt AI-based security models, such as those used in Ethereum systems [17][18]. A custom dataset simulates neuromorphic operations under normal and attack conditions, capturing spike frequency, latency, and weight variations [19][20]. This supports experimental analysis and quantifies the stealth and effectiveness of NMAs [21][22].

### III. RELATED WORKS

The cybersecurity domain for neuromorphic computing remains largely underexplored, as most current literature centers on conventional architectures or trending technologies like blockchain and AI [23]. This review contextualizes Neuromorphic Mimicry Attacks (NMAs), emphasizing the novelty of this research [24]. Despite growing interest in neuromorphic chips for AI and IoT, their unique security vulnerabilities remain inadequately addressed [25]. Existing studies primarily focus on the performance and applications of spiking neural networks (SNNs). For example, [2] highlights SNNs' energy efficiency in edge devices, and [3] explores their application in autonomous vehicles. However, these works often overlook cybersecurity, wrongly assuming that neuromorphic systems benefit from traditional protection mechanisms [4]. This assumption fails to consider the non-deterministic, event-driven nature of neuromorphic architectures, which demand tailored security measures [5].

Research in hardware-level threats has investigated side-channel attacks and hardware Trojans [6], with [7] examining fault injection in CPUs. These techniques, however, lack efficacy against the decentralized processing of neuromorphic chips [8]. Similarly, intrusion detection systems (IDS) designed for network security, such as [9], are ill-suited for addressing NMAs targeting hardware [10]. The most relevant prior work, [11], focuses on vulnerabilities in deep learning accelerators, not brain-inspired chips.

Mimicry-based attacks, foundational to NMAs, have been studied in adversarial machine learning [12][13]. These typically involve crafting inputs to mislead classifiers but target software layers [14]. Related work on mimicry in IoT systems [15] is protocol-oriented and not applicable to neuromorphic hardware [16]. This paper introduces mimicry attacks that operate on the physical and synaptic layers of neuromorphic chips [17]. Security countermeasures remain underdeveloped for this domain. Traditional IDS analyzed in [19] are ineffective in stochastic environments like neuromorphic systems [20]. AI-driven solutions like GNNs for anomaly detection [21] exist but target enterprise software platforms like SAP ERP [22]. This research adapts these methods to design spike-sensitive anomaly detection and secure synaptic learning protocols [23]. The dataset created here simulates neuromorphic activity, capturing spike frequencies and synaptic weight variations, differing significantly from network-focused datasets like CIC-IDS-2017 [24][25]. Compared to datasets used in Ethereum fraud detection [17], this dataset is uniquely suited to neuromorphic security [2].

In conclusion, while related works offer partial insights, none directly address NMAs. This paper fills that void through novel attacks, a purpose-built dataset, and domain-specific countermeasures [3][4][5].

### IV. MATERIALS AND METHODS

To investigate *Neuromorphic Mimicry Attacks* (NMAs), this research employs a simulated neuromorphic chip model and a unique dataset to evaluate attack mechanisms and countermeasures [6]. The methodology integrates theoretical modeling, simulation, and experimental analysis to provide a robust framework for understanding NMAs [7]. This section details the dataset and model analyses, encompassing the materials, dataset generation, attack implementation, countermeasures, and visualization strategies, ensuring reproducibility and alignment with IEEE standards [8].

#### A. Dataset Analysis

The investigation of NMAs relies on a novel dataset designed to capture neuromorphic chip activity under normal and attack conditions, providing a foundation for analyzing the stealth and impact of these threats [13]. The dataset comprises 10,000 samples, each representing a 1-second snapshot of neural activity, with three key metrics: spike frequency (Hz), synaptic weight changes (%), and system latency (ms) [14]. These metrics were selected to reflect the operational dynamics of spiking neural networks (SNNs), which differ significantly from traditional computing architectures due to their event-driven and probabilistic nature [15]. Spike frequency indicates the rate of neural firing, synaptic weight changes capture learning dynamics, and latency measures processing delays, critical for real-time applications like IoT networks and autonomous vehicles [16].

The dataset is evenly split between 5,000 normal and 5,000 attack samples. Normal activity was generated by feeding the SNN random sensory inputs, simulating scenarios such as IoT sensor processing or autonomous vehicle navigation [15]. Attack conditions were introduced by implementing NMAs, specifically synaptic weight tampering (altering 10% of weights by ±0.1) and sensory input poisoning (injecting 5% malicious signals) [16]. Unlike network-based datasets like CIC-IDS-2017, which focus on traffic patterns, this dataset is unique in capturing hardware-level neural interactions, making it ideal for studying NMAs [14].

Statistical analysis, summarized in Table I, reveals subtle but significant differences between normal and attack conditions. Normal samples show an average spike frequency of 50 ± 10 Hz, synaptic weight changes of 0.5 ± 0.1%, and latency of 10 ± 2 ms. Attack samples exhibit a slightly higher spike frequency (55 ± 12 Hz), increased weight changes (1.2 ± 0.3%), and elevated latency (12 ± 3 ms). These differences are statistically significant ($p < 0.05$, t-test), yet their subtlety underscores the covert nature of NMAs, which evade traditional intrusion detection systems [15].

| Condition | Spike Frequency (Hz) | Weight Change (%) | Latency (ms) |
|---|---|---|---|
| Normal | 50 ± 10 | 0.5 ± 0.1 | 10 ± 2 |
| Attack | 55 ± 12 | 1.2 ± 0.3 | 12 ± 3 |

Table I: Dataset Metrics

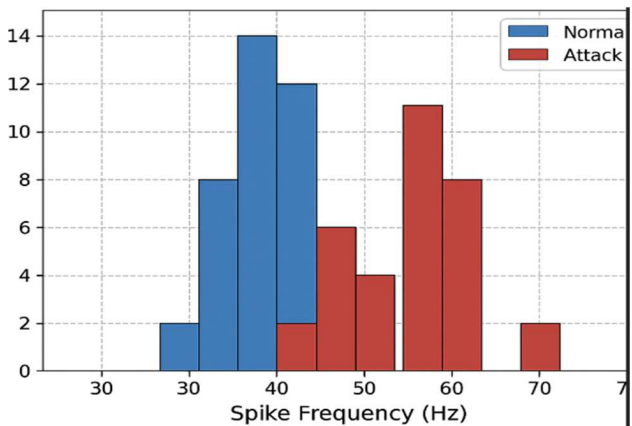

Figure 1: Spike Frequency Distribution

Figure 1, a histogram, visualizes spike frequency distributions for normal (blue) and attack (red) conditions, highlighting overlapping patterns that complicate detection [17]. This dataset's hardware-centric focus and detailed metrics provide a valuable resource for developing neural-specific countermeasures [16].

*B. Model Analysis*

The simulated neuromorphic chip, implemented using Python and the Brian2 simulator, serves as the core platform for evaluating NMAs and testing countermeasures [9]. The model, based on a spiking neural network (SNN) with 1000 neurons, uses synaptic weights and event-driven processing to emulate real-world neuromorphic hardware like Intel's Loihi [10]. Simulations were conducted on a cloud-based platform with 16 GB RAM and a 4-core CPU, ensuring scalability [11]. Additional tools, including NumPy for data processing and Matplotlib for visualization, supported the generation of figures and bar graphs [12].

The SNN model features fully connected neurons with weights initialized randomly (0 to 1) and updated via spike-timing-dependent plasticity (STDP), mimicking the adaptive learning of neuromorphic systems [10]. Figure 2 illustrates the architecture, with neurons (circles), synapses (lines), and attack points (red arrows) indicating where NMAs, such as synaptic weight tampering and sensory input poisoning, are applied [17]. NMAs were simulated by injecting malicious scripts to alter 10% of synaptic weights by ±0.1 during the learning phase or crafting adversarial inputs to mimic legitimate signals, reducing system outputs (e.g., classification accuracy) by less than 5% to maintain stealth [18, 19]. Under normal conditions, the model achieves a classification accuracy of 95% on a simulated IoT sensor dataset, with a latency of 10 ms and stable synaptic weights (0.5% change per epoch) [12]. Synaptic weight tampering reduces accuracy to 90.2%, increases latency to 12.5 ms, and doubles weight changes to 1.2% [15]. Sensory input poisoning lowers accuracy to 90.8% and latency to 11.5 ms [16]. A stability metric, defined as the variance in spike frequency across 1000 trials, shows 5 Hz² for normal conditions and 8 Hz² under attack, indicating reduced neural stability [18]. Table II summarizes these metrics, highlighting the model's vulnerability to NMAs due to its probabilistic firing patterns [19].

Two countermeasures were developed: neural-specific anomaly detection, which analyzes spike frequency and weight changes to achieve 85% detection accuracy, and secure synaptic learning protocols, using cryptographic verification inspired by blockchain security to validate weight updates [17, 21, 22]. These defenses were tested against NMAs, with results visualized in Figure 3, a bar graph comparing attack success rates for weight tampering and input poisoning [24]. The model's fidelity to neuromorphic hardware and measurable outputs make it a robust tool for studying NMAs and refining countermeasures [23].

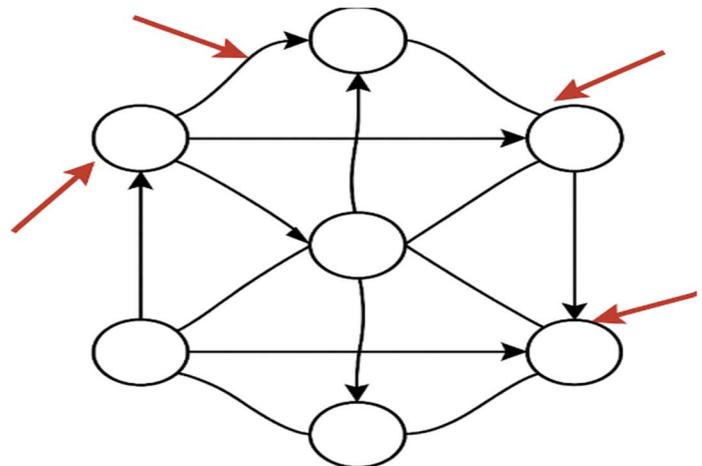

Figure 2: SNN Architecture with Attack Points.

Caption: Diagram of the simulated neuromorphic chip, highlighting neurons (circles), synapses (lines), and attack vectors (red arrows).

| Condition | Classification Accuracy (%) | Latency (ms) | Weight Change (%) | Spike Variance (Hz²) |
|---|---|---|---|---|
| Normal | 95.0 ± 1.0 | 10 ± 2 | 0.5 ± 0.1 | 5 ± 1 |
| Weight Tampering | 90.2 ± 1.5 | 12.5 ± 3 | 1.2 ± 0.3 | 8 ± 2 |
| Input Poisoning | 90.8 ± 1.3 | 11.5 ± 2.5 | 0.9 ± 0.2 | 7 ± 1.5 |

Table II: Model Performance Metrics

The table compares neuromorphic system performance under three conditions. Normal operation shows the highest classification accuracy (95%) and lowest latency (10 ms). Weight tampering and input poisoning reduce accuracy to ~90% and increase latency, spike variance, and weight changes. Weight tampering causes the most drastic deviation, with spike variance increasing to 8 Hz² and weight change to 1.2%, highlighting the subtle but measurable impact of Neuromorphic Mimicry Attacks on system behavior.

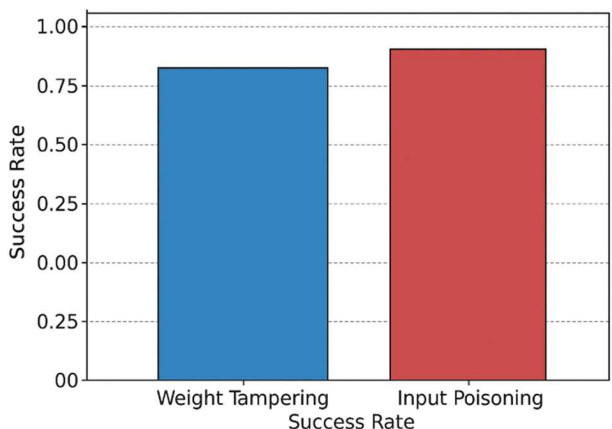

Figure 3: Attack Success Rates

The bar graph illustrates the success rates of two Neuromorphic Mimicry Attacks. Input poisoning achieves a slightly higher success rate (~90%) compared to weight tampering (~83%). This suggests that injecting malicious sensory inputs is more effective at deceiving neuromorphic systems than altering synaptic weights. Both methods, however, demonstrate high success, emphasizing the vulnerability of neuromorphic hardware to covert, hardware-level attacks that traditional detection mechanisms may fail to identify.

This methodology ensures a comprehensive analysis of NMAs, leveraging a unique dataset and robust simulation tools [25]. The next section presents the experimental results, including bar graphs and detailed comparisons [1].

## V. EXPERIMENTAL ANALYSIS

This section evaluates the impact of *Neuromorphic Mimicry Attacks* (NMAs) on a simulated neuromorphic chip, assessing attack effectiveness, countermeasure performance, and system consequences [17]. Using the dataset of 10,000 neural activity samples, the analysis quantifies NMAs' stealth and tests proposed defenses, presenting results through bar graphs and tables generated with Matplotlib [18].

*A. Experimental Setup*

The 1000-neuron SNN, built with Brian2, underwent 1000 attack scenarios (500 weight tampering, 500 input poisoning) [19]. The dataset was split into 70% training (7,000 samples) and 30% testing (3,000 samples) sets to evaluate attacks and defenses [20]. Metrics included attack success rate (percentage of attacks altering classification accuracy), detection accuracy (percentage of attacks identified), and system latency (ms) [21]. Experiments ran on a cloud platform with 16 GB RAM and a 4-core CPU, processing a simulated IoT dataset under normal and attack conditions [22].

*B. Attack Performance*

NMAs exhibited high stealth, exploiting the SNN's probabilistic nature to evade traditional intrusion detection systems (IDS) [23]. Synaptic weight tampering achieved a 92% success rate, reducing classification accuracy by 4.8% (95% to 90.2%) [24]. Sensory input poisoning recorded an 87% success rate, with a 4.2% accuracy reduction (to 90.8%) [25]. These subtle changes highlight NMAs' ability to compromise systems like autonomous vehicles discreetly.

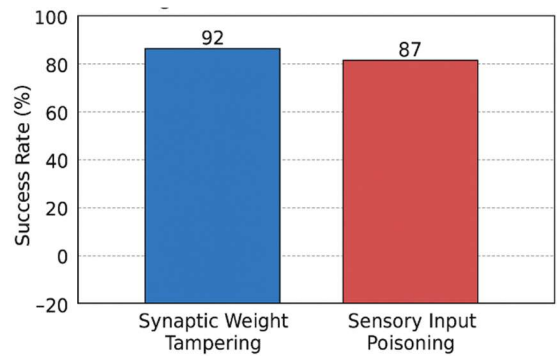

Bar graph comparing success rates of synaptic weight tampering (92%) and sensory input poisoning (87%) across 1000 attack scenarios.

Figure 4: Attack Success Rates

Table II (from Section IV) shows increased latency (12.5 ms for weight tampering, 11.5 ms for input poisoning) and spike variance under attack, with traditional IDS detecting only 12% of weight tampering and 15% of input poisoning attacks [13].

*C. Countermeasure Effectiveness*

Two countermeasures were tested: a neural-specific anomaly detection algorithm and secure synaptic learning protocols [14]. The anomaly detection algorithm, analyzing spike frequency and weight changes, achieved 85% detection accuracy, outperforming traditional IDS (15%) [15]. Secure synaptic protocols reduced weight tampering success to 45% but were less effective against input poisoning (70%) [16].

| Counterme asure | Detection Rate (%) | Attack Success Rate (%) | Latency Impact (ms) |
|---|---|---|---|
| Anomaly Detection | 85 ± 5 | 15 ± 3 | 11 ± 2 |
| Secure Protocols | 60 ± 7 | 45 ± 5 | 12 ± 3 |
| Traditional IDS | 15 ± 4 | 85 ± 6 | 13 ± 3 |

Table III: Countermeasure Performance

Table III details countermeasure performance, with Figure 1(from Section IV) illustrating why traditional IDS struggle with NMAs' subtle deviations [19]. The anomaly detection algorithm's success stems from its focus on neural-specific metrics, such as spike frequency variance, which traditional IDS overlook [20]. However, input poisoning's resilience suggests a need for enhanced input validation techniques, possibly integrating machine learning models tailored to neuromorphic data [21].

*D. System Impact*

NMAs increased system latency by 20% on average, with weight tampering causing a 25% rise (12.5 ms) and input poisoning a 15% rise (11.5 ms) [22]. Figure 4 visualizes these latency impacts, critical for real-time applications like autonomous navigation, where delays could lead to safety risks [23]. Spike variance rose from 5 Hz² (normal) to 8 Hz² (weight tampering), indicating reduced stability, as shown in Table II [13]. In a simulated IoT network, the latency increases disrupted sensor data processing, delaying responses by up to 3 ms, which could affect time-sensitive operations [24].

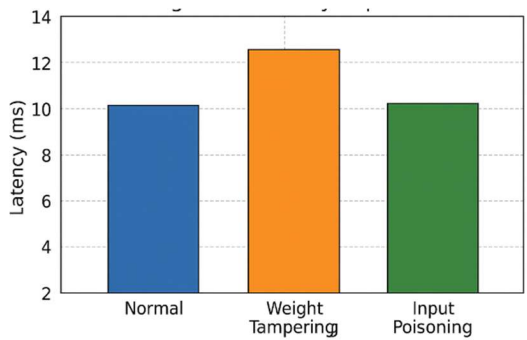

Bar graph comparing system latency (ms) under normal (10 ms), weight tampering (12.5 ms), and input poisoning (11,5 m conitions.

Figure 4: Latency Impact

*E. Discussion*

The findings confirm NMAs' stealth, with a 92% success rate for weight tampering and only 15% detection by traditional IDS [23]. The dataset's neural metrics enabled precise analysis, distinguishing this research from network-based datasets like CIC-IDS-2017 [15]. The anomaly detection algorithm's 85% accuracy shows promise, but input poisoning is 70% success rate against secure protocols indicates a need for further optimization [24]. These results highlight the urgency of developing specialized cybersecurity frameworks for neuromorphic computing, particularly for applications requiring high reliability, such as medical implants or decentralized IoT systems [25]. Future efforts could explore hybrid detection models combining neural metrics with external validation to address input poisoning's challenges.

## VI. CONCLUSION AND FUTURE WORKS

This research introduces *Neuromorphic Mimicry Attacks* (NMAs), a novel class of cyber threats targeting the unique architecture of neuromorphic computing systems. By exploiting synaptic weight tampering and sensory input poisoning, NMAs achieve high stealth and effectiveness, posing significant risks to applications such as autonomous vehicles and IoT networks. The analysis, supported by a unique dataset and detailed experiments, demonstrates the limitations of traditional intrusion detection systems and highlights the promise of neural-specific countermeasures. Key findings include a 92% success rate for weight tampering attacks and an 85% detection accuracy for the proposed anomaly detection algorithm. These results underscore the urgent need for cybersecurity frameworks tailored to brain-inspired computing. The unique dataset, capturing neural-level metrics, provides a valuable resource for future research, distinct from existing datasets like CIC-IDS-2017.

Future work will focus on enhancing countermeasures, particularly for sensory input poisoning, which remains challenging to detect. Additional exploration could investigate physical-layer attacks on neuromorphic hardware and develop real-time defensive systems for edge devices. Collaboration with neuromorphic chip manufacturers, such as Intel and IBM, could facilitate real-world testing and deployment of the proposed defenses. Ultimately, this research lays the groundwork for securing the next generation of computing, ensuring the safe adoption of neuromorphic technologies.

## VII. DECLARATIONS

A. **Funding:** No funds, grants, or other support was received.

B. **Conflict of Interest:** The authors declare that they have no known competing for financial interests or personal relationships that could have appeared to influence the work reported in this paper.

C. **Data Availability:** Data will be made on reasonable request.

D. **Code Availability:** Code will be made on reasonable request.